\documentstyle[12pt]{article}

\large
\newcommand{\be}{\begin{eqnarray}}
\newcommand{\ee}{\end{eqnarray}}
\newcommand\del{\partial}

\begin{document}
\setlength{\baselineskip}{21pt}
\pagestyle{empty}
\vfill
\eject
\begin{flushright}
SUNY-NTG-94/19
\end{flushright}

\vskip 2.0cm
\centerline{\bf Random Matrix Theory and QCD$_3$}
\vskip 2.0 cm
\centerline{J.J.M. Verbaarschot and I. Zahed}
\vskip .2cm
\centerline{Department of Physics}
\centerline{SUNY, Stony Brook, New York 11794}
\vskip 2cm

\centerline{\bf Abstract}

We suggest that the spectral properties near zero virtuality
of three dimensional QCD, follow from a Hermitean random matrix model.
The exact spectral density is derived for this family of random matrix models
both for even and odd number of fermions. New sum rules for the inverse powers
of the eigenvalues of the Dirac operator are obtained. The issue of anomalies
in random matrix theories is discussed.

\vfill
\noindent
\begin{flushleft}
SUNY-NTG-94/19\\
April  1994
\end{flushleft}
\eject
\pagestyle{plain}

In four dimensions, the spontaneous breakdown of chiral symmetry in QCD
is characterized by the order parameter $<\overline{q}q>$ that relates to
the spectral density of the QCD Dirac operator near zero virtuality, but
many level spacings away from the origin \cite{BANKS-CASHER-1980}.
In order to analyze the
dynamics of the order parameter, it is therefore natural to study not
only the asymptotic limit of the spectrum near zero virtuality, but also
the way this limit is approached as the thermodynamic limit is taken.
More concretely, we will analyze the spectrum near zero virtuality at
finite space-time volume.

The spectrum of the Dirac operator fluctuates over the ensemble of
gauge fields. This raises the question whether these fluctuations
may be independent of the particular dynamics of the system. As we know
from the study of quantum chaos and universal conductance fluctuations
and from experimentally measured spectra of
compound nuclei \cite{BOHIGAS-GIANNONI-1984}, correlations between
levels of
the order of several level spacing are universal. Because of this
they can be described by a random matrix model
which has only the $symmetries$ of the system as input.
Our conjecture is that the same is true for the correlations of
the eigenvalues of the Dirac operator near zero virtuality
\cite{SV-1993,VZ-1993,JJV-1994}.

Since chirality is not defined in odd space-time dimensions,
the issue of spontaneous breaking is more subtle in three dimensions.
However, we may still consider the possibility of a spontaneous
breaking of  global flavor symmetry and parity. Three dimensional QCD
($QCD_3$) maybe of interest to four dimensional QCD at high temperature
\cite{HOT} and to quantum antiferromagetism \cite{WIEG}.
It may also be used to describe certain disordered condensed matter
systems \cite{FRADKIN-1985}.

In this letter we would like to argue that the spontaneous breaking of
flavor and/or parity in three dimensional QCD is related to the behavior of
the spectral density near zero virtuality. Using the universality arguments
presented above, we will derive the so called microscopic spectral density
\cite{SV-1993,VZ-1993},
from random matrix theory. However, in order to formulate the correct
model we have to analyze the symmetries of the underlying theory. The outcome
will be a Hermitean random matrix model which satisfies
the 3-dimensional analogues of the four dimensional Leutwyler-Smilga sum
rules \cite{LS}. The way the anomaly shows up in random matrix theories will
be briefly discussed.

Consider $QCD_3$ with $N_f$ flavors,
\be
{\cal L} = -\frac 14 {\rm Tr} F^2 +
\sum_{a=1}^{N_f} \overline{q}_a\left( \hat D + m_a \right) q_a
\label{1}
\ee
where $\hat D= \gamma_\mu D_\mu$ is the covariant derivative. The
$q_a$ are two component spinors in the fundamental representation of
$SU(N_c)$
Color and spin indices have been suppressed.
In the limit where all the masses
are equal $m_a = m$, there is a global $U(N_f)$ symmetry of which the $U(1)$
part relates to the baryon number, and is conserved independent of $m_a$.
In three dimensions,
the gamma matrices are Pauli matrices. There is no distinction
between left and right handed fermions so that $U(N_f)$
does not have the chiral structure of four dimensions.

In three dimensions, the parity operation is a reflection about one axis,
say $x_1$. This operation is implemented by $\gamma_0\gamma_2$ on the
spinors. As a result $\overline{q}q\rightarrow -\overline{q}q$
under parity. For massless quarks, (\ref{1}) is
invariant under parity. However, the Dirac operator for a given color field
is not. The total symmetry group of
(\ref{1}) is
$SU(N_c)\times U(N_f)\times Z_2$ at  the classical level.
If the quark masses are arranged in pairs of opposite signs but equal
magnitude, then (\ref{1}) is still parity preserving. Indeed, for $N_f$ even,
a rearrangement of (\ref{1}) gives
\be
{\cal L} = -\frac 14 {\rm Tr} F^2 +
\sum_{a=1}^{N_f} \overline{q}_a\rlap/ D q_a +
m \left(\sum_{a=1}^{N_f/2} \overline{q}_aq_a -\sum_{b=1}^{N_f/2}
\overline{q}_bq_b\right)
\label{2}
\ee
which is $SU(N_c)\times \left( U(N_f/2)\times U(N_f/2)\right)\times Z_2$
at the tree level. Here, the discrete $Z_2$ is the product of the usual
parity (${\bf P}_1$)
with the interchange $a\rightarrow b$ (${\bf E}$).

At the quantum level, global anomalies can crop in. For an odd number of
flavors, parity is either broken explicitly by quark masses, or radiatively
in the massless limit, with the appearance of a Chern-Simons term. For an even
number of flavors a rearrangement of the masses in doublets as in (\ref{2})
causes the anomaly to vanish. In this case, parity is a good symmetry even at
the quantum level.
The parity broken phase is a screening phase, with free triality. The quarks
are heavy and carry fractional statistics. This phase is related to the
flux phase of quantum antiferromagnets as discussed by Wiegman
\cite{WIEG}. The parity
symmetric phase maybe confining, with some analogy with four dimensional QCD.

Could it be that flavor and/or parity symmetry
are spontaneously broken in $QCD_3$?
In so far, there is no lattice simulation to support
that\footnote{However, lattice calculation have been performed for
$QED_3$ \cite{SACHA},
and a nonzero value of $\langle \bar q q \rangle$
was found.}. We suspect that
for an even number of flavors $U(N_f)\times Z_2$ breaks spontaneously to
$U(N_f/2)\times U(N_f/2)\times Z_2$
while for an odd number of flavors $Z_2$
is radiatively broken by the anomaly in the massless case, and explicitly
in the massive case. By a generalization of the Vafa-Witten theorem
\cite{VAFA-WITTEN-1984} to $QCD_3$ it follows that the absolute values of all
condensates are equal. As follows from the derivation of the Banks-Casher
relation, the quark mass and the condensate for each flavor have the same
sign if the symmetry is broken $spontaneously$.
Therefore, we can introduce the order parameter
\be
|\Sigma| = \frac 1{N_f}\sum_{a=1}^{N_f} |<\overline{q}_a q_a>|.
\label{4}
\ee

Throughout, we will study the spectrum of the Dirac operator at finite
space-time volume. In the chiral limit the spectrum
is well defined and contains information on the structure of the vacuum,
including the way the flavor symmetry is broken. In the large $N_c$ limit,
a rerun of the Coleman-Witten
argument \cite{COLEMAN} for $QCD_3$ suggests that in the even case the
symmetry is broken according to
$U(N_f)\rightarrow U(N_f/2)\times U(N_f/2)$. This
is confirmed below using an effective Lagrangian method in the saddle point
approximation. In this spirit and for even $N_f$, the condensates will be
combined in pairs of opposite sign.

To understand the generic behaviour
of the Dirac spectrum near zero virtuality, we note that the massless Euclidean
Dirac operator $i \hat D [A] $ in an arbitrary background field is
Hermitean and does not commute with ${\bf P_1E}$. By analogy with four
dimensions, the random matrix model with the symmetries of (\ref{2}) is
\be
Z(m)= \int {\cal D}T P(T) \prod_{a=1}^{N_f}
{\rm det} (iT+m_a),\label{6}
\ee
where the Haar measure is over $N\times N$ hermitean matrices, and the $m_j$
are the eigenvalues of the mass matrix.
The weight distribution $P(T)$ is chosen to be Gaussian consistent with
no additional input but the symmetries of the system.
We note that the fermion determinants make the integrand not necessarily
positive. This ensemble with the determinant replaced
by its absolute value is also known as the generalized gaussian ensemble
\cite{NAGAO-SLEVIN-1993}.

The order parameter (\ref{4}) follows from
\be
\Sigma = - {\lim_{m_a\to0}}{\lim_{N\to\infty}} \frac 1N \frac d{dm_a}
{\rm log} \,Z
\label{7}
\ee
If we were to define the continuum "spectral density" by
\be
\rho_C (\lambda ) = {\lim_{m \to 0}}{\lim_{N \to\infty}}
\langle \frac 1N \sum_n \delta (\lambda -\lambda_n)\rangle
\label{8}
\ee
where the expectation value is over the partition function (\ref{2}),
then (\ref{7}) can be rewritten as
\be
\Sigma = i\pi \rho_C (0) - {\bf P}\int d\lambda
\frac {{\rho_C}(\lambda )}{\lambda}
\label{9}
\ee
For even spectra, the principle value part vanishes.

The partition function (\ref{6}) can be evaluated if we recall that the
general decomposition for Hermitian matrices is $T= U\Lambda U^\dagger$, where
$\Lambda$ is a diagonal matrix. Using the eigenvalues and eigenangles of
$T$ as new integration variables, the partition function can be rewritten as
\be
Z(m) = \int \prod_k d\lambda_k \prod_{k<l} |\lambda_k -\lambda_l|^2
\prod_k\prod_{a=1}^{N_f}(i\lambda_k +m_a)
\exp(-\frac{N \Sigma^2}2 \sum_k
\lambda_k^2)
\label{10}
\ee
The one-point function $\rho(\lambda)$ is defined as the integral over all
eigenvalues in this partition function except one. Its normalization can
be expressed through $Z = \int d\lambda\rho(\lambda)$.
{}From (\ref{10}) it follows immediately that
$\rho(-\lambda)  = (-1)^{N N_f} \rho(\lambda)$.
For an even number of flavors $\rho(\lambda)$ is positive in the chiral limit
and can be interpreted as the spectral density of the Dirac operator.
For an odd number of flavors $\rho(\lambda)$ is still an even function for
even $N$ but is not necessarily positive definite. For odd $N$ and an odd
number of flavors it is an odd
function. In this case $Z(0) = 0$ and
$Z(m) \sim \left . \del_m Z \right |_{m=0}$.
In the  calculation of the condensate (see (\ref{7})) the derivative of the
partition function cancels and to leading order in $m$ we obtain
$\Sigma = \lim_{m\rightarrow 0}\lim_{N\rightarrow \infty}(-1)/Nm =0$.
Thus, chiral symmetry remains unbroken in this limit.

Before discussing the spectral density related to
(\ref{6}), we will derive
the finite volume partition function in the static limit, as for
$QCD_4$ \cite{SV-1993}. This is achieved by rewriting the fermion determinant
as Grassmann integrals, and averaging over $T$. The result is a four-fermion
interaction. The effective partition function is obtained
after bosonisation and a saddle point approximation.
To leading order in $1/N$ only the saddle point with the eigenvalues
of $\sigma$ in opposite pairs contributes \cite{VWZ}. The resulting
partition function has a 'hyperbolic' symmetry \cite{VWZ} and is given by
\be
Z({\cal M}) = \int_{U\in SU(N_f)} DU \exp(N\Sigma
{\rm Tr}{\cal  M}\, U I U^{\dagger}),
\label{eff7}
\ee
where we have extended the integral over the coset $U(N_f)/U(N_f/2)\times
U(N_f/2)$ to $SU(N_f)$, and
$I= {\rm diag}({\bf 1}_{N_f/2}, -{\bf 1}_{N_f/2})$.
Equivalently, the result (\ref{eff7}) could also
be arrived at using general symmetry arguments \cite{LS}.

The linear term in ${\cal M}$ in
the expansion of $Z({\cal M})$ vanishes.
The integrals over $SU(N_f)$ that occur in the terms of ${\cal
O}({\cal M}^2)$ are well known. We find
\be
Z({\cal M}) = Z(0)\left[1 + \frac {\Sigma^2 N^2 }2
\frac {N_f^2}{N_f^2 -1} \left( \frac 1{N_f} {\rm Tr} {\cal M}^2 -
\frac 1{N_f^2} {\rm Tr}^2 {\cal M} \right )  + \cdots\right ].
\label{eff9}
\ee
On the other hand, the $QCD_3$ partition function
can be expanded as
\be
Z(m) = Z(0) \langle \left(
1 + i {\rm Tr} {\cal M} \sum_k \frac 1{\lambda_k} +
\frac 12 ({\rm Tr} {\cal M}^2 - {\rm Tr}^2 {\cal M}) \sum_{k} \frac
1{\lambda_k^2} - \frac 12 {\rm Tr}^2 {\cal M} \sum_{k\ne l} \frac
1{\lambda_k \lambda_l} + \cdots \right) \rangle,
\label{eff10}
\ee
resulting in the sum rules
\be
\frac 1{N^2}\sum_{k} \frac 1{\lambda_k^2} = \frac{N_f}{N_f^2 -1}
\Sigma^2
\qquad{\rm and}\qquad
\frac 1{N^2}\sum_{k\ne l} \frac 1{\lambda_k \lambda_l} = -\frac{\Sigma^2}{N_f +
1}.
\label{eff11}
\ee
Note that the average over the second term in (\ref{eff10}) vanishes,
in agreement with the effective partition function.

For odd $N_f$ it is not possible to organize the saddle points in opposite
pairs. Because of the Jacobian and to leading order in $1/N$ the saddle points
occur at $(N_f + 1)/2$ eigenvalues with $\pm 1/\Sigma$ and $(N_f-1)/2$
eigenvalues with $\mp 1/\Sigma$. Both sets of saddle points
cannot be transformed
into each other by a unitary transformation. However, all saddle point in one
set are connected by a unitary transformation. The sum of the two saddle points
leads to the partition function
\be
Z({\cal M}) = \int_{U\in SU(N_f)} DU \cosh(N\Sigma {\rm Tr}{\cal M}\,
  U I U^{\dagger}), \label{eff12a},
\ee
for even $N$, and ${\rm cosh}\rightarrow {\rm sinh}$ for odd $N$.
The matrix $I$ is now ${\rm diag}({\bf 1}_{(N_f+1)/2},
-{\bf 1}_{(N_f-1)/2})$.
The finite volume partition function is therefore based on the coset
$U(N_f)/U((N_f-1)/2)\times U((N_f+1)/2)$.
Sum rules follow from comparison to the $QCD_3$ partition function
\be
\frac 1{N^2}\sum_{\lambda_k} \frac 1{\lambda_k^2} = \frac{\Sigma^2}{N_f}
\qquad{\rm and}\qquad
\frac 1{N^2}\sum_{k\ne l} \frac 1{\lambda_k \lambda_l} = -\frac{\Sigma^2}{N_f}
\label{eff16}
\ee
for even values of $N$.

With these results in mind, we now turn to the evaluation of the spectral
density using random matrix theory. The microscopic spectral density
and its correlation functions will
be the master formulae for all sum rules. The construction of the spectral
density, can be achieved with the help of the orthogonal
polynomial method from random matrix theory
(see for example \cite{MEHTA-1991}). For even $N_f$ we proceed as
for the Gaussian Unitary Ensemble\cite{VZ-1993}.
The result for the spectral density can be written as
\be
\rho(\lambda) = \sum_{k=0}^{N-1} \frac 1{r_k}P_k(\lambda)P_k(\lambda)
\lambda^{N_f}  \exp(-a^2 \lambda^2),
\label{17}
\ee
where the $P_k$ are orthogonal polynomials that satisfy
\be
\int_{-\infty}^{\infty} d\lambda P_k(\lambda) P_l(\lambda) \lambda^{N_f}
\exp(-a^2 \lambda^2)= r_k \delta_{kl},
\label{18}
\ee
and $r_k$ is a normalization factor.
For even $N_f$ these polynomials
can be expressed in terms of the generalized Laguerre polynomials
\be
 P_{2k}(\lambda) = L_{k}^{\frac{N_f-1}{2}}(a^2\lambda^2 )
\qquad{\rm and}\qquad
 P_{2k+1}(\lambda) = \lambda L_{k}^{\frac{N_f+1}{2}}(a^2\lambda^2 ).
\label{19}
\ee
Using (\ref{18}) and (\ref{19}) the normalization factors are determined
readily.

The spectral density follows immediately from (\ref{17}).
It can be written as the sum of two terms.
Both sums can be evaluated with the
help of the Christoffel-Darboux formula. The result is straightforward,
and yields an exact analytical expression for the level density.
Its microscopic limit
\be
\rho_S(z) = \lim_{N\rightarrow\infty} \frac 1N \rho(\frac zN).
\label{21}
\ee
follows from the asymptotic properties of the generalized Laguerre
polynomials. The result for even $N_f$ is \cite{NAGAO-SLEVIN-1993}
\be
\rho_S(z) = \frac {\Sigma^2 z}4 \left [
J_{\frac{N_f -1}2}^2(z) - J_{\frac{N_f +1}2}(z)J_{\frac{N_f -3}2}(z)
+
J_{\frac{N_f +1}2}^2(z) - J_{\frac{N_f +3}2}(z)J_{\frac{N_f -1}2}(z)\right ].
\label{22}
\ee
It reproduces the first sum rule of (\ref{eff11}).

For $N_f $ $odd$ it is not possible to construct
a set of orthogonal polynomials
satisfying (\ref{18}). For example, the zeroth order polynomial cannot be
orthogonal to the first order polynomial. Therefore, in the Vandermonde
determinant, we replace the $2k-1$'th row
by the difference of the $2k$'th and the
$2k-1$'th row and the $2k$'th row by the sum of the $2k-1$'th and the
$2k$'th row. The polynomials in the $2k-1$'th and the $2k$'th row therefore
have the same degree. In this case the level density is again given by
(\ref{17})
but the orthogonal polynomials $P_{2k-1}$ and $P_{2k}$ have the same degree.
The construction of these polynomials is straightforward
\be
P_{2k-1} =  (1-\lambda a)
L_{k}^{\frac {N_f}2}(\lambda^2 a^2)\qquad{\rm and}\qquad
P_{2k} =  (1+\lambda a)
L_{k}^{\frac {N_f}2}(\lambda^2 a^2).
\label{24}
\ee
Repeating the steps leading (\ref{22}) we arrive at the microscopic spectral
density
\be
\rho_S(z) = (-1)^{N/2}\frac {\Sigma^2 z}2
\left( J_{N_f/2}^2(z) - J_{N_f/2 +1}(z)J_{N_f/2 -1}(z)\right) ,
\label{27}
\ee
which reproduces the first spectral sum rule of (\ref{eff16}).
The case of odd $N$ and odd $N_f$ will be discussed elsewhere.

Finally, we would like to comment on the issue of the anomaly.
In three dimensions, ${\rm det}(i \hat D_3[A])$ is
noninvariant under an adiabatic switch-on or -off
of a large gauge transformation \cite{REDLICH}. The gauge noninvariance is
usually followed by a spectral flow in the fermionic spectrum with level
crossing at zero. This means an overall change in the sign of the fermion
determinant. Thus the anomaly and the phase of the fermion determinant are
related. Indeed, consider the case when one eigenvalue
say $\lambda$, crosses 0. The relevant factor in our case that determines
the change in phase is given by
$\prod_{a=1}^{N_f} (\lambda + im_a)$.
When the crossing is completed with $|m_a| \ll |\lambda|$, the
determinant acquires the extra phase
\be
\exp(\pm i\pi \sum_{a=1}^{N_f} {\rm sgn}\, m_a).
\label{29b}
\ee
This result is to be contrasted with the continuum result \cite{REDLICH}
\be
\exp({\pm i\pi \sum_{a=1}^{N_f} {\rm sgn}\, m_a \,\,{\bf W}_{cs}[A]}),
\label{30}
\ee
where ${\bf W}_{cs}[A]$ is the Chern-Simons action in Euclidean space.
The imaginary character of (\ref{30}) in Euclidean space,
follows from the fact that ${\bf W}_{cs}$ is $T$-odd. The sign
ambiguity in our case, corresponds to the sign ambiguity left by the
Pauli-Villars regulator in the continuum. Thus, the net sign effect
in random matrix theory amounts to a Chern-Simons term.

To conclude: we  have argued that on the basis of universality
arguments that the spectral density of $QCD_3$ near
zero virtuality follows from a hermitean random matrix model. We have
explicitly constructed the microscopic spectral densities for an even and odd
number of flavors. The resulting sum rules are in agreement with the expected
results following from an effective Lagrangian formulation based solely
on symmetry arguments. For an even number of flavors, we have suggested using
the effective Lagrangian formulation, that $U(N_f)$ is likely to be maximally
broken to $U(N_f/2)\times U(N_f/2)$. We have also shown that the
Chern-Simons term
has a natural explanation in the context of random matrix theory.

Finally, we want to note that the random matrix model used in this work
is based on the assumption that flavor symmetry is broken spontaneously.
It is this
assumption that leads to the finite volume static partition function
as quoted in the text.
Since, in general, odd dimensions do not sustain semiclassical physics,
this leads us to the questions: what is the mechanism behind the spontaneous
breaking of flavor symmetry? Could it be that this mechanism is
also responsible for the spontaneous breaking of chiral symmetry in four
dimensions ? These points deserve further investigation.

\vglue 0.6cm
{\bf \noindent  Acknowledgements \hfil}
\vglue 0.4cm
This work was supported in part  by the US DOE grant DE-FG-88ER40388.

\setlength{\baselineskip}{15pt}


\end{document}